\documentclass[12pt]{article}
\usepackage{authblk}
\usepackage[bookmarksnumbered, colorlinks, plainpages]{hyperref}
\usepackage{amsmath, amsthm, amscd, amsfonts, amssymb, graphicx, color, booktabs}
\DeclareUnicodeCharacter{03B3}{\ensuremath{\gamma}}

\numberwithin{equation}{section}
\definecolor{email}{rgb}{0.00,0.00,0.84}
\begin{document}
\setcounter{page}{1}

\title{\large \bf 12th Workshop on the CKM Unitarity Triangle\\ Santiago de Compostela, 18-22 September 2023 \\ \vspace{0.3cm}
\LARGE \textit{CP} violation measurements in three-body charmless \textit{B} decays}


\author{Lucas Falcão\textsuperscript{1}\footnote{lucas.falcao@cern.ch} on behalf of the LHCb Collaboration \\
        \textsuperscript{1}Centro Brasileiro de Pesquisas Físicas, Rio de Janeiro, Brazil \\ }

\maketitle

\begin{abstract}

This study reports \textit{CP} asymmetry measurements in charmless three-body decays of \textit{B} mesons, both using the Run II data from proton-proton collisions at a center-of-mass energy of 13 $TeV$, collected by the LHCb detector from 2015 to 2018, with an integrated luminosity of 5.9 $fb^{-1}$. Significant \textit{CP} asymmetries are observed in $B^\pm \rightarrow \pi^\pm \pi^+ \pi^- $ and $B^\pm \rightarrow K^\pm K^+ K^- $ decays, while the previously observed asymmetry in $B^\pm \rightarrow \pi^\pm K^+ K^- $ decays is confirmed, and the \textit{CP} asymmetry of $B^\pm \rightarrow K^\pm \pi^+ \pi^- $ decays is found to be compatible with zero \cite{Direct_CPV}. Additionally, a new method is used to measure the \textit{CP} asymmetry in charmless $B \rightarrow PV$ decays, showing significant asymmetry in $B^\pm \rightarrow K^\pm \pi^+ \pi^- $ decays dominated by $\rho(770)$, representing the first observation of \textit{CP} violation in this process. Other decay channels show \textit{CP} asymmetries compatible with zero \cite{BtoPV}.

\end{abstract} \maketitle

\section{Direct \texorpdfstring{\textit{CP}}{CP} violation in charmless three-body decays of \texorpdfstring{$B^\pm$}{} mesons}

The decays of heavy mesons into three-body can be comprehended by considering a combination of intermediate states featuring resonances. Within this framework, the \textit{CP}-integrated asymmetries ($A_{CP}$) observed in charmless three-body decays of \textit{B} mesons are the sum of asymmetries associated with each intermediate state, weighted by their respective fractions. This study investigates \textit{CP} violation in four charmless \textit{B} meson decays involving three charged pseudoscalar particles: $B^\pm \rightarrow K^\pm \pi^+ \pi^- $,  $B^\pm \rightarrow K^\pm K^+ K^- $, $B^\pm \rightarrow \pi^\pm K^+ K^- $, and $B^\pm \rightarrow \pi^\pm \pi^+ \pi^- $. The analysis focuses on measuring phase-space integrated \textit{CP} asymmetries and asymmetries within specific regions of the Dalitz plots. To take into account the production asymmetries of the \textit{B} mesons, the decay $B^\pm \rightarrow J/\psi(\rightarrow \mu^+ \mu^-) K^\pm$ is employed as a control channel.

\begin{figure} [h!]
  \begin{center}
    \includegraphics[height=190pt,width=400pt]{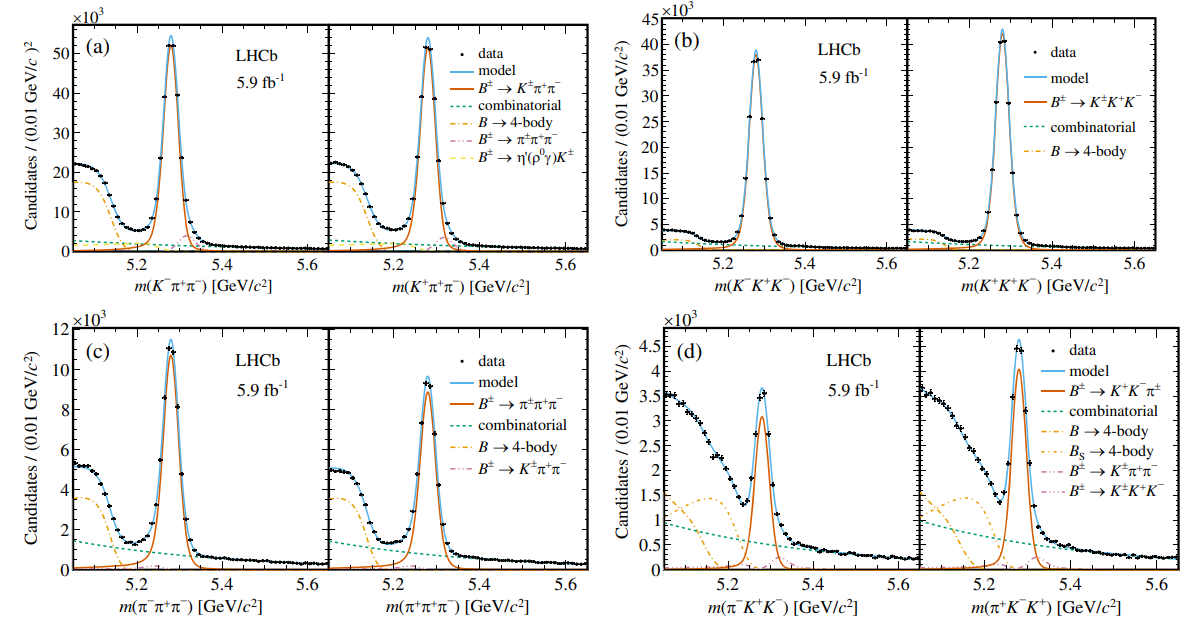}
    \caption{Invariant mass spectra of the four decays in the range [5050–5650] MeV/$c^2$.}
    \label{fig:invariant_mass_fits}
  \end{center}
\end{figure}

The yield and raw asymmetry of each decay are derived through simultaneous unbinned extended maximum likelihood fits to the $B^\pm$ invariant mass distributions, as shown in Fig. \ref{fig:invariant_mass_fits}. Signal components within each of the four channels are modeled by a combination of a Gaussian and two Crystal Ball functions, sharing a common mean. Combinatorial backgrounds are parametrized by an exponential function. Backgrounds arising from partially reconstructed four-body \textit{B} decays are represented by the ARGUS function \cite{Argus} convolved with a Gaussian resolution function. Simulated event samples are used to model detector efficiency effects. The resultant signal yields and raw asymmetries ($A_{raw}$) are outlined in Table \ref{tab:yields}.

\begin{table}[h]
\centering
\caption{Signal yields, asymmetries, and efficiency ratio R of the four decays.}
\begin{tabular}{cccc}
\hline
Decay mode & Total yield & $A_{raw}$ & $R = \langle\epsilon^-\rangle/\langle \epsilon^+ \rangle$   \\ \hline
$B^\pm \rightarrow K^\pm \pi^+ \pi^-$ & $499200 \pm 900$ & $+0.006 \pm 0.002$ & $1.0038 \pm 0.0027$ \\
$B^\pm \rightarrow K^\pm K^+ K^-$ & $365000 \pm 1000$ & $-0.052 \pm 0.002$ & $0.9846 \pm 0.0024$ \\
$B^\pm \rightarrow \pi^\pm \pi^+ \pi^-$ & $101000 \pm 500$ & $+0.090 \pm 0.004$ & $1.0354 \pm 0.0037$ \\
$B^\pm \rightarrow \pi^\pm K^+ K^-$ & $32470 \pm 300$ & $-0.132 \pm 0.007$ & $0.9777 \pm 0.0032$ \\
\hline
\end{tabular}
\label{tab:yields}
\end{table}

\subsection{Phase-space integrated and localized \textit{CP} asymmetries}

Phase-space integrated \textit{CP} asymmetries are derived by correcting the raw asymmetry for selection efficiency effects to obtain the efficiency-corrected raw asymmetries ($A^{corr}_{raw}$), and considering the production asymmetry ($A_P$). The physical \textit{CP} asymmetry can be expressed in terms of $A^{corr}_{raw}$ and $A_P$:

\begin{equation}
    A_{CP} = \frac{A^{corr}_{raw} - A_P }{1 - A^{corr}_{raw}A_P}.
\end{equation}

\noindent The \textit{CP} asymmetries of the four $B^\pm \rightarrow h^\pm h'^+ h'^-$
modes are measured to be: 

\begin{gather}
    A_{CP}(B^\pm \rightarrow K^\pm \pi^+ \pi^-) = +0.011 \pm 0.002 \pm 0.003 \pm 0.0003, \\ \nonumber
    A_{CP}(B^\pm \rightarrow K^\pm K^+ K^-) = -0.037 \pm 0.002 \pm 0.002 \pm 0.003, \\ \nonumber
    A_{CP}(B^\pm \rightarrow \pi^\pm \pi^+ \pi^-) = +0.080 \pm 0.004 \pm 0.003 \pm 0.003, \\ \nonumber
    A_{CP}(B^\pm \rightarrow \pi^\pm K^+ K^-) = -0.114 \pm 0.007 \pm 0.003 \pm 0.003. \nonumber
\end{gather}

The first uncertainty is statistical, followed by systematic uncertainty, with a third component stemming from the limited knowledge of the \textit{CP} asymmetry in the control channel \cite{Direct_CPV}. The computation of significance of \textit{CP} asymmetries is 8.5$\sigma$ for $B^\pm \rightarrow K^\pm K^+ K^-$ decays, 14.1$\sigma$ for $B^\pm \rightarrow \pi^\pm \pi^+ \pi^-$ decays, and 13.6$\sigma$ for $B^\pm \rightarrow \pi^\pm K^+ K^-$ decays, marking the first observation of \textit{CP} asymmetries in these decay channels. In the case of $B^\pm \rightarrow K^\pm \pi^+ \pi^-$ decays, the \textit{CP} asymmetry significance is 2.4$\sigma$, aligning with the expectation of \textit{CP} conservation.

The so-called Dalitz plot represents directly the decay dynamics, allowing the examination of various resonant and non-resonant components to scrutinize \textit{CP} asymmetry effects. To visualize localized asymmetries, the $A_{CP}$ is calculated in bins of the phase space \cite{Mirandizing}. 

\begin{figure}[htbp]
  \centering
  \begin{minipage}[b]{0.49\textwidth}
    \centering
    \includegraphics[width=6cm]{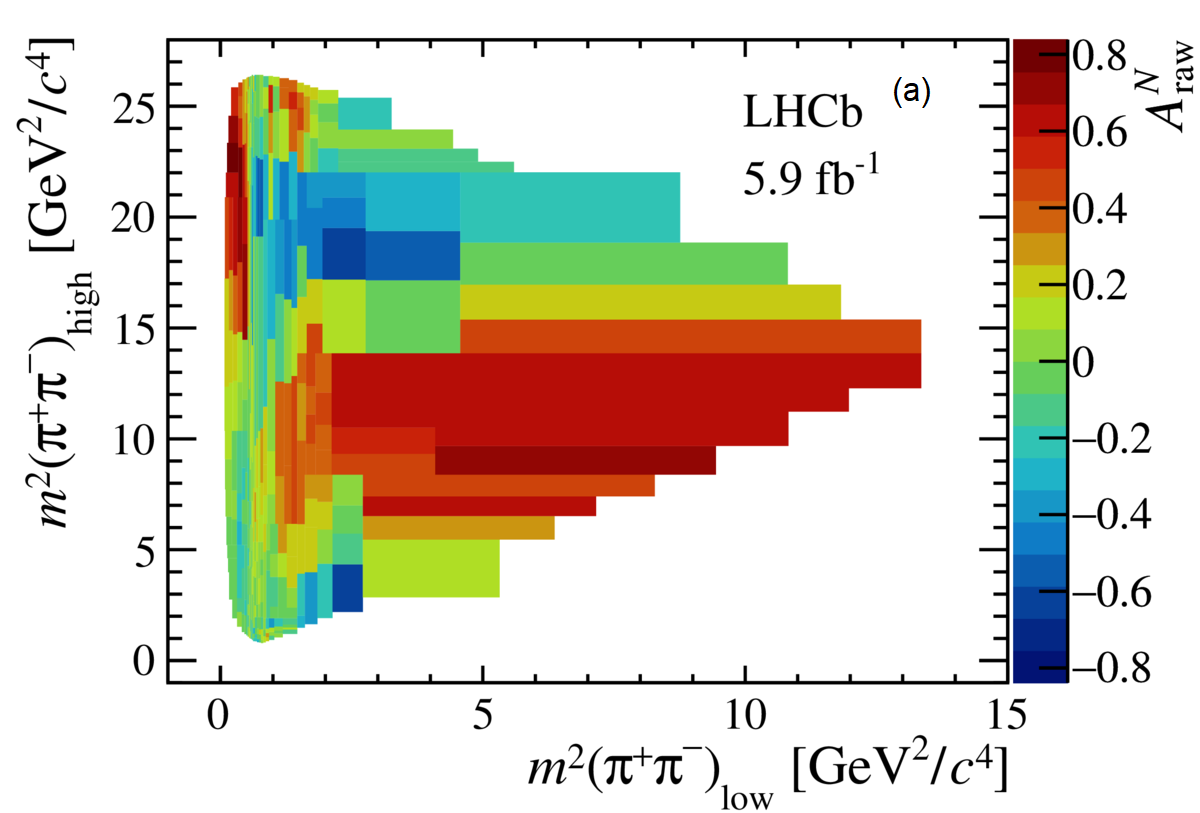}
  \end{minipage}
  \hfill
  \begin{minipage}[b]{0.49\textwidth}
    \centering
    \includegraphics[width=6cm]{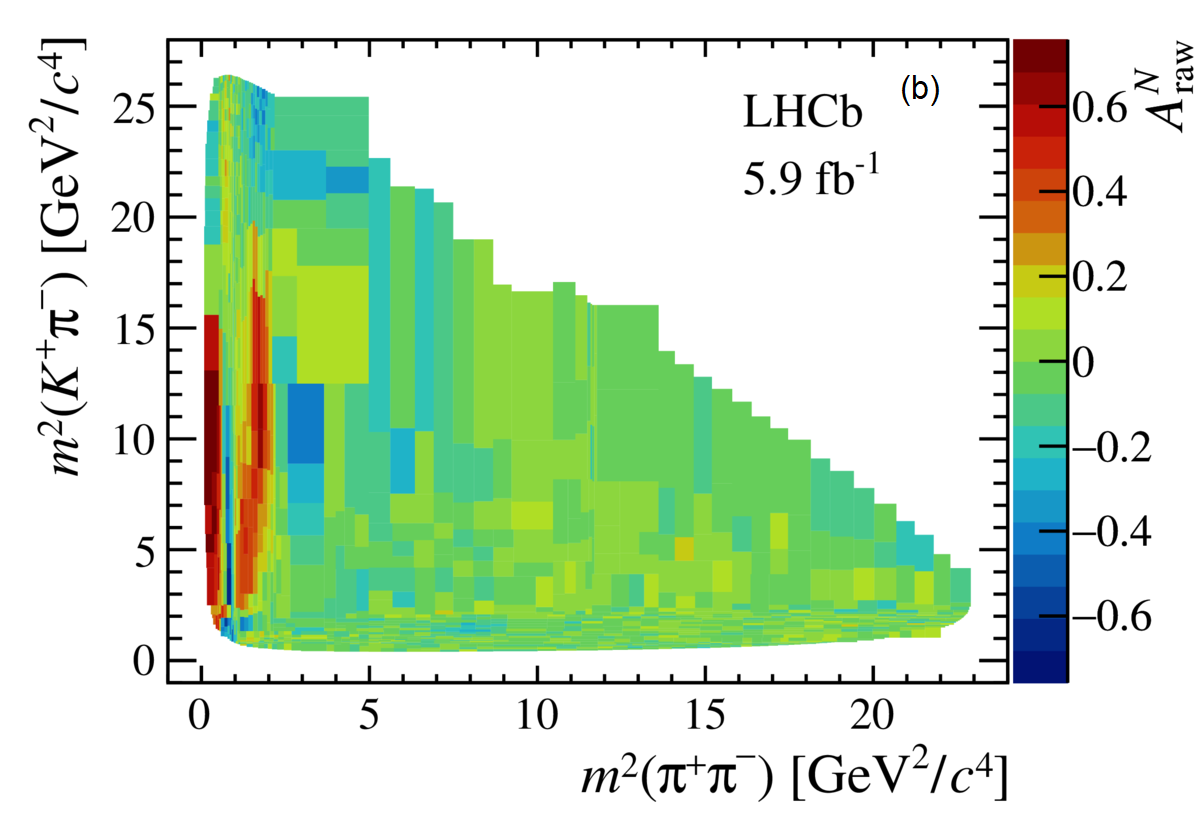}
  \end{minipage}
  
  \begin{minipage}[b]{0.49\textwidth}
    \centering
    \includegraphics[width=6cm]{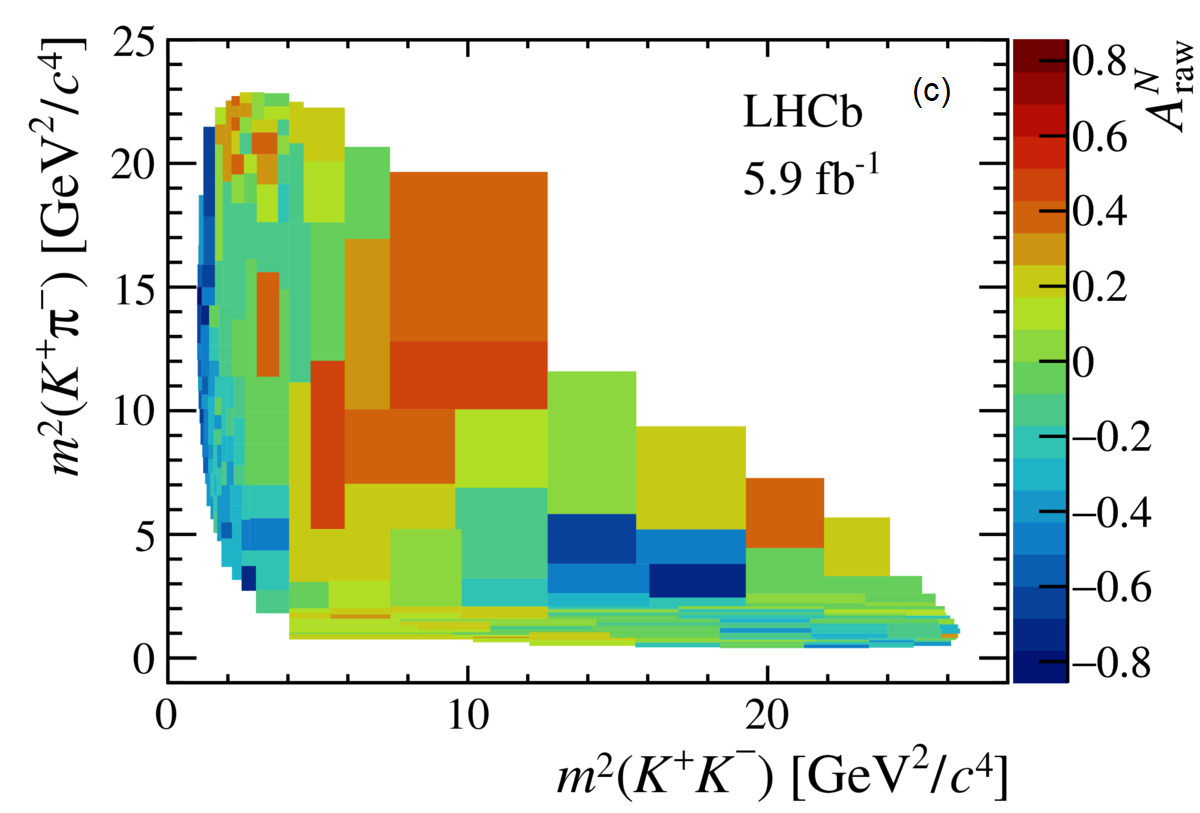}
  \end{minipage}
  \hfill
  \begin{minipage}[b]{0.49\textwidth}
    \centering
    \includegraphics[width=6cm]{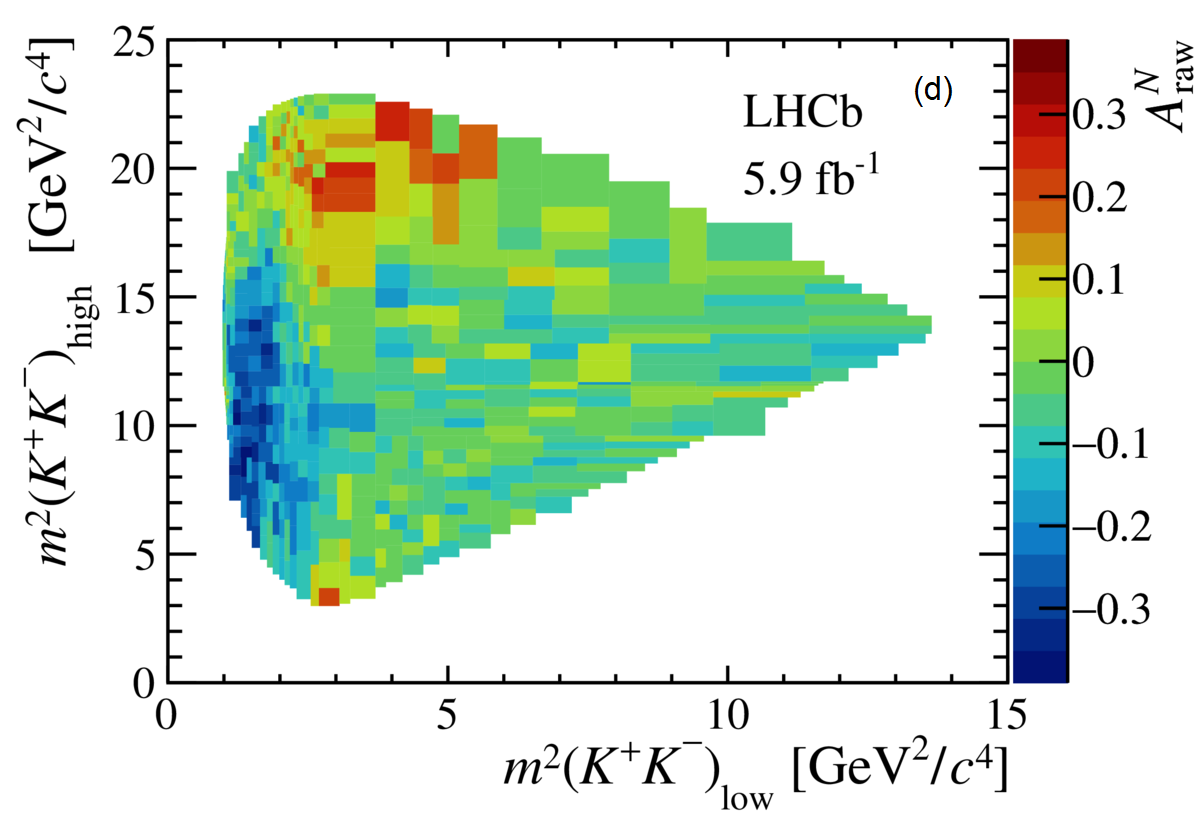}
  \end{minipage}

    \caption{Asymmetry distribution in bins of the Dalitz plot for (a) $B^\pm \rightarrow \pi^\pm \pi^+ \pi^-$; (b) $B^\pm \rightarrow K^\pm \pi^+ \pi^-$; (c) $B^\pm \rightarrow \pi^\pm K^+ K^-$; and (d) $B^\pm \rightarrow K^\pm K^+ K^-$.}
    \label{fig:mirandizing}
\end{figure}

For $B^\pm \rightarrow K^\pm \pi^+ \pi^-$, the $m^2(K^+ \pi^-)$ axis reveals a change in the asymmetry sign, see (b) in Figure \ref{fig:mirandizing}. For the rescattering region of $B^\pm \rightarrow \pi^\pm K^+ K^-$, one can see a nearly constant \textit{CP} asymmetry along the $m^2(K^+ \pi^-)$ axis, see (c). The $B^\pm \rightarrow K^\pm K^+ K^-$ decays also reveal CP
asymmetries in the $m^2(K^+ K^-)$ projections in the
rescattering region, and another
change in the asymmetry sign is observed, in an opposite
direction with respect to $B^\pm \rightarrow K^\pm \pi^+ \pi^-$, see (d). A large asymmetry is observed for the $B^\pm \rightarrow \pi^\pm \pi^+ \pi^-$, where there is an indication of $\chi_c^{0}(1P)$ contribution around 11.6 Ge$V^2/c^4$, see (a).

\section{Search for direct \textit{CP} violation in charged charmless \texorpdfstring{$B \to PV$}{B to PV} decays}

This study investigates quasi-two-body \textit{B} meson decays into PV (pseudoescalar and vector) final states, leading to three-body configurations arising from the subsequent decays of the vector meson. 

A novel approach, independent of a full amplitude analysis, is employed. The method is based on three
key features of three-body \textit{B} decays \cite{BtoPV}: the large phase space;
the dominance of scalar and vector resonances with masses
below or around 1 GeV/$c^2$; and the signatures
of the resonant amplitudes in the Dalitz plot.



Usually, the decay amplitudes of $B^\pm$ are represented as a coherent sum of intermediate amplitudes, where the magnitude and phase of each amplitude are adjustable parameters. When a single vector resonance interferes with a scalar component, the decay amplitudes can be expressed as $ \mathcal{M}_\pm = a_\pm^V e^{i\delta^V_\pm}F_V^{BW}cos\theta(s_\perp, s_\parallel) + a_\pm^S e^{i\delta^S_\pm}F_S^{BW}$. The Matrix element squared is:

\begin{equation}
\begin{aligned}
    |\mathcal{M}_\pm|^2 = (a^V_\pm)^2(cos\theta)^2|F_V^{BW}|^2 + (a^S_\pm)^2|F_S^{BW}|^2 + 2a^V_\pm a^S_\pm cos \theta |F_V^{BW}|^2 |F_S^{BW}|^2 \\
    \times \big\{ cos(\delta^V_\pm - \delta^S_\pm) [(m^2_V - s_\parallel)(m^2_S - s_\parallel) + (m_V\Gamma_V)(m_S\Gamma_S)] \quad  \quad  \quad\\
    + sin(\delta^V_\pm - \delta_\pm^S)[(m_S\Gamma_S)(m^2_V - s_\parallel) - (m_V\Gamma_V)(m^2_S - s_\parallel)] \big\}, \quad \quad \quad
    \label{eq:matrix_element}
\end{aligned}
\end{equation}

\noindent where $a^V_\pm$ and $\delta^V_\pm$ are the magnitude and phase of the vector and $a^S_\pm$  and $\delta^S_\pm$ are the magnitude and phase of the scalar resonances. The $\theta(s_\perp, s_\parallel)$ is the helicity angle. For the low mass and sufficiently narrow resonances, a parabolic dependence of $cos\theta$ only on $s_\perp$ is a good approximation. Assuming that the phases and amplitudes do not depend on $s_\perp$, the Eq.\ref{eq:matrix_element} can be reduced to a quadratic polynomial in $cos\theta(m^2_V,s_\perp)$ as:

\begin{equation}
    |\mathcal{M}_\pm|^2 = f(cos\theta(m^2_V,s_\perp)) = p_0^\pm + p_1^\pm cos\theta(m_V^2,s_\perp) +p_2^\pm cos^{2}\theta(m_V^2,_\perp)
    \label{eq:matrix_element_2}
\end{equation}

Given that the decay rates are proportional to $|\mathcal{M}_\pm|^2$, the \textit{CP} asymmetry $A_{CP}^{V}$ in $B\rightarrow PV$ is given as function of $p_2^\pm$:

\begin{equation}
    A_{CP}^{V} = \frac{|\mathcal{M}_{-}|^2 - |\mathcal{M_{+}}|^2}{|\mathcal{M_{-}}|^2 + |\mathcal{M_{+}}|^2} = \frac{p_2^- - p_2^+}{p_2^- + p_2^+},
    \label{eq:polynomial}
\end{equation}

Given the approximation $cos\theta(s_\parallel,s_\perp)\approx cos\theta(m^2_V,s_\perp)$, $cos\theta$ becomes a linear function of $s_\perp$ \cite{kajantie}, with this approximation the \textit{CP} asymmetry can be obtained from the distribution of $s_\perp$. The function on Eq. \ref{eq:polynomial} is used to fit the histograms of data projected onto $s_\perp$ axes in order to determine the $p_{0,1,2}^\pm$ and then calculate the $A_{CP}^{V}$.

\begin{figure}[htbp]
  \centering
  \begin{minipage}[b]{0.49\textwidth}
    \centering
    \includegraphics[width=6cm]{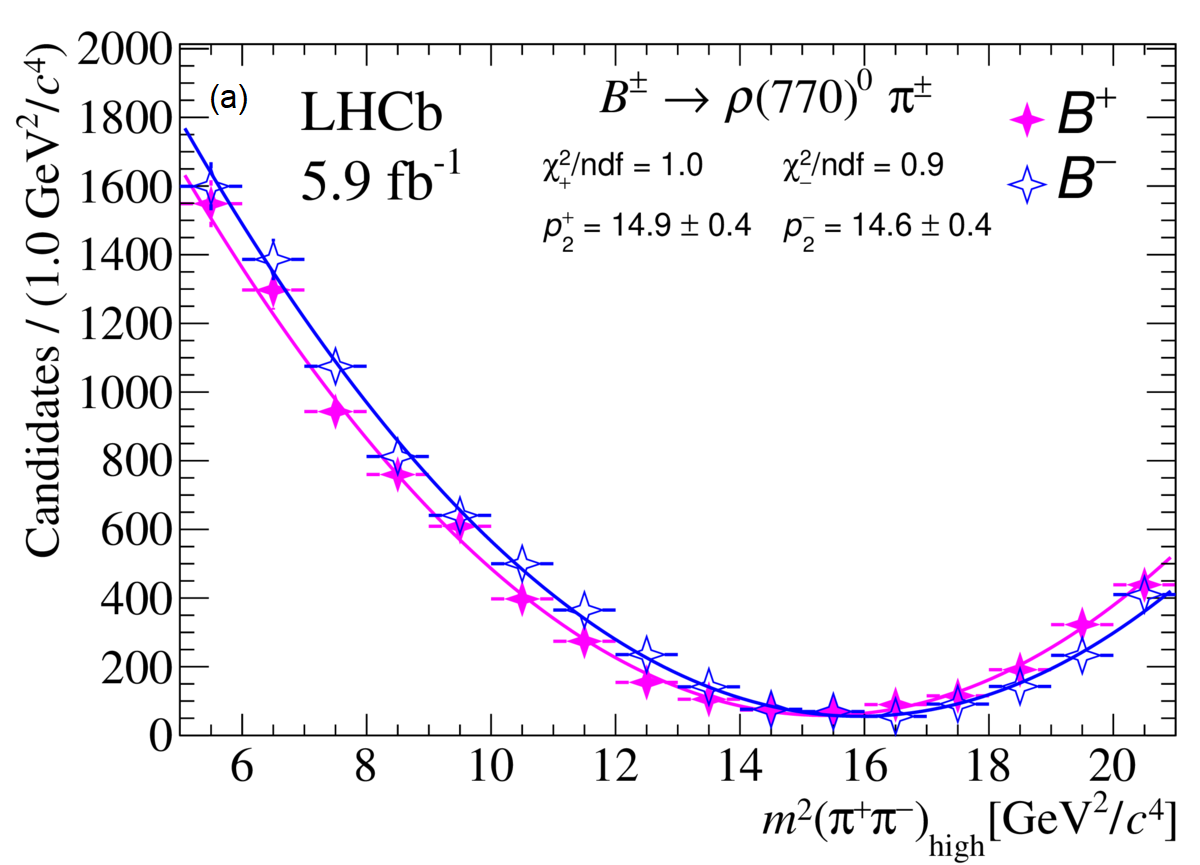}
  \end{minipage}
  \hfill
  \begin{minipage}[b]{0.49\textwidth}
    \centering
    \includegraphics[width=6cm]{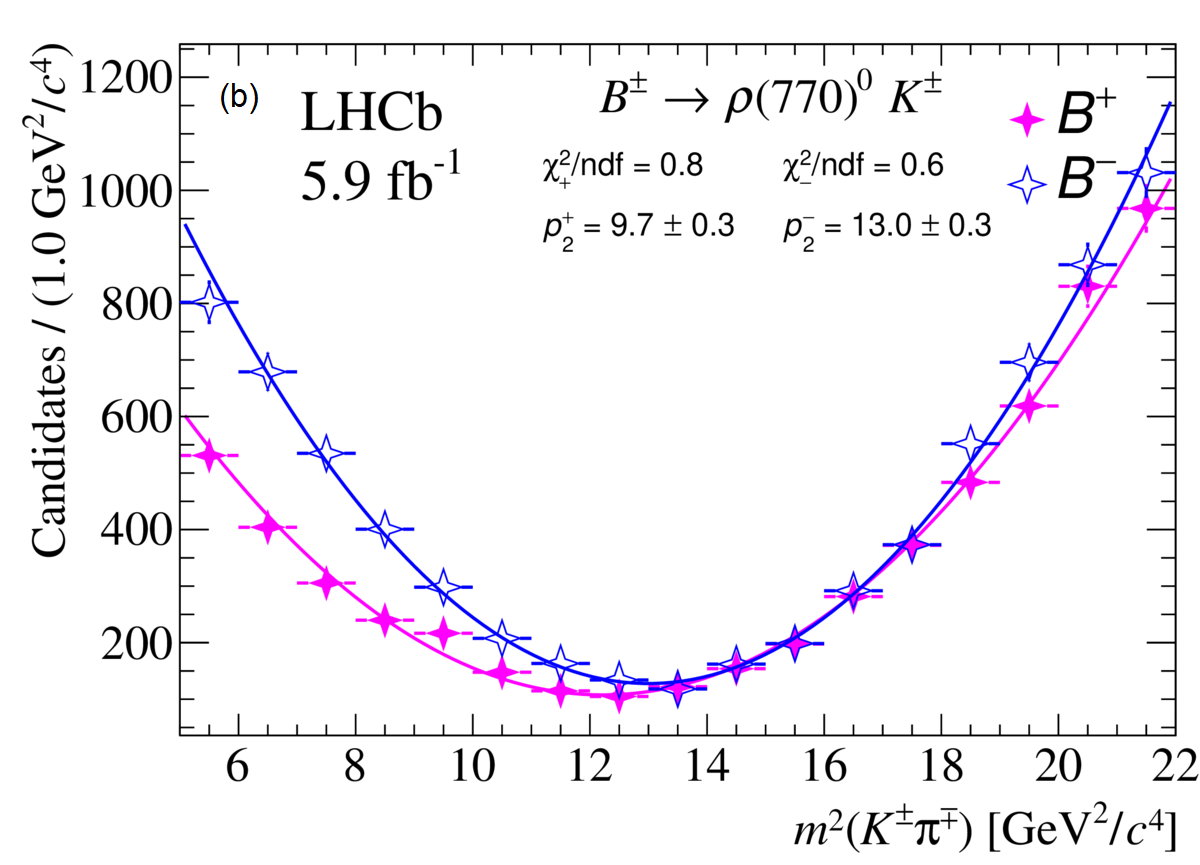}
  \end{minipage}
  
  \begin{minipage}[b]{0.49\textwidth}
    \centering
    \includegraphics[width=6cm]{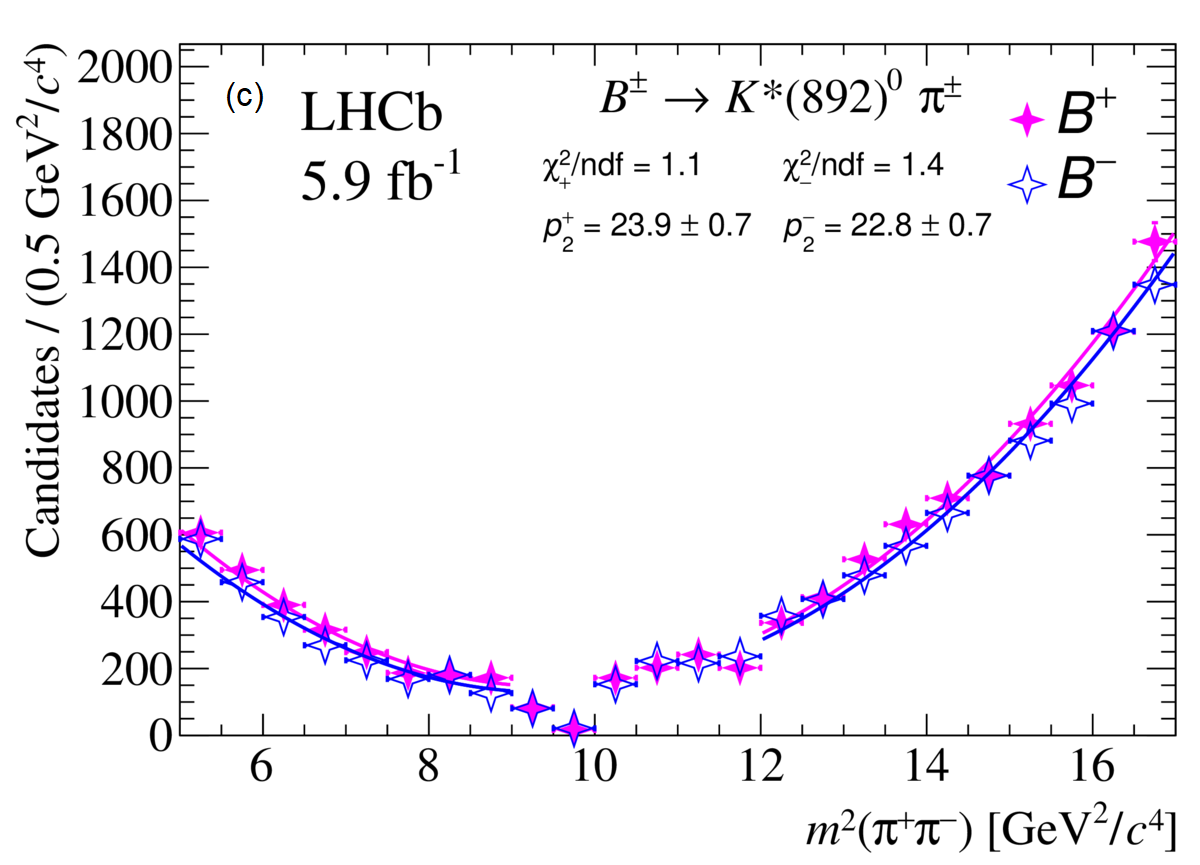}
  \end{minipage}
  \hfill
  \begin{minipage}[b]{0.49\textwidth}
    \centering
    \includegraphics[width=6cm]{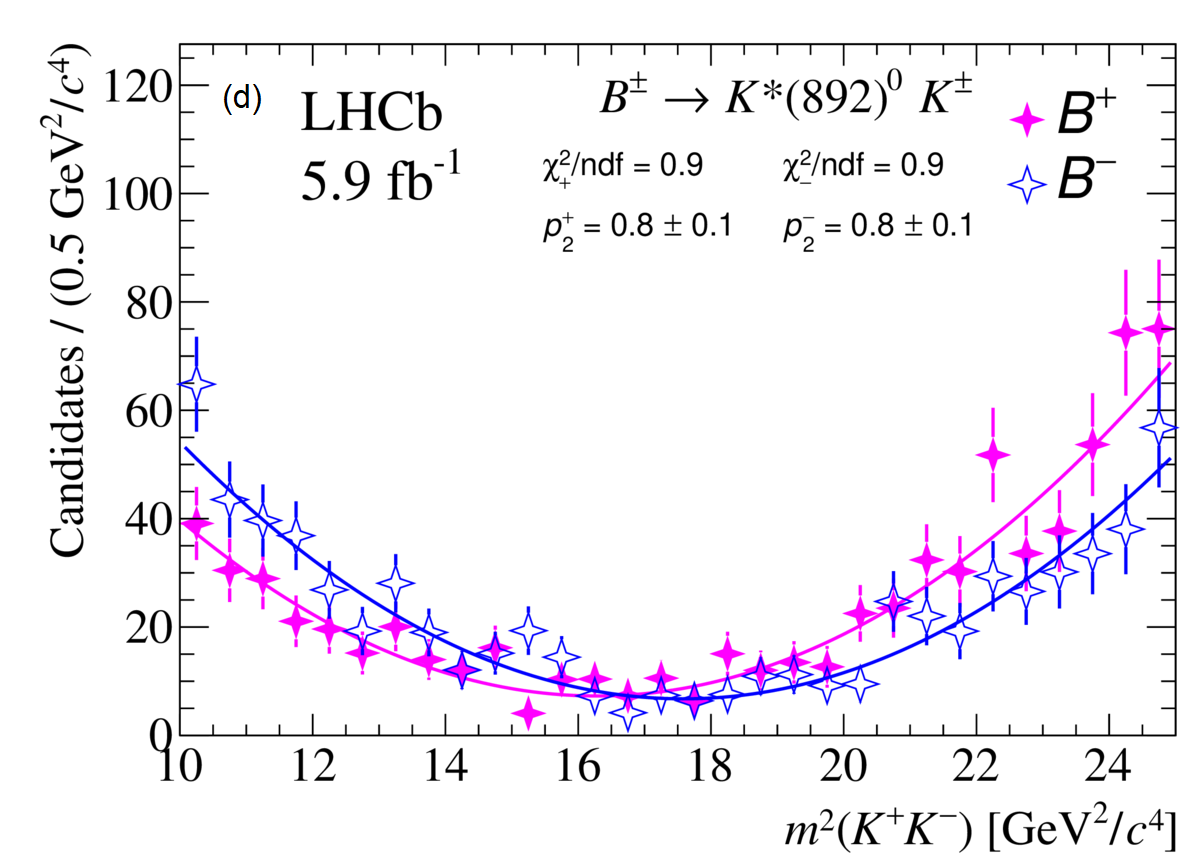}
  \end{minipage}
  
  \begin{minipage}[b]{0.49\textwidth}
    \centering
    \includegraphics[width=6cm]{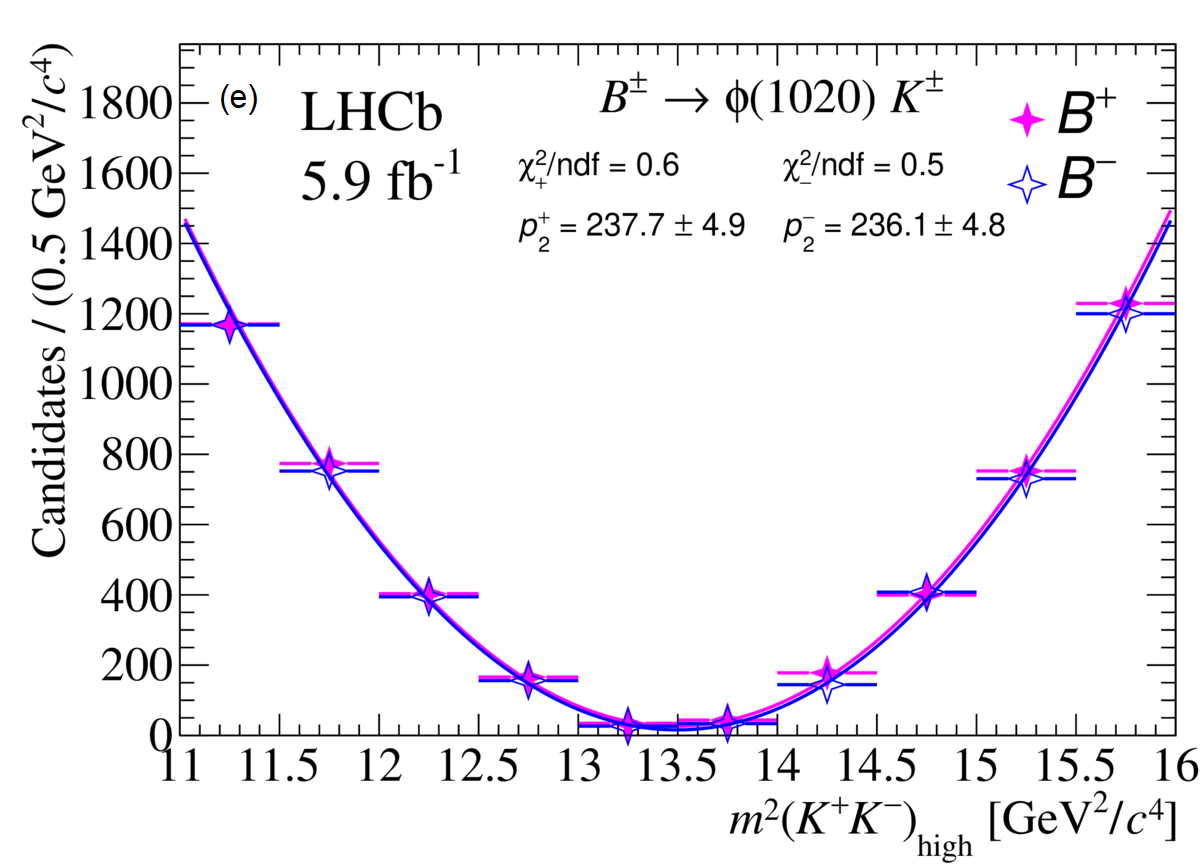}
  \end{minipage}

  \caption{Distribution of $s_\perp$ for $B^\pm$ and the corresponding quadratic fits for (a) $\rho(770)$ in $B^\pm \rightarrow \pi^\pm \pi^+ \pi^-$, (b) $\rho(770)$ in $B^\pm \rightarrow K^\pm \pi^+ \pi^-$, (c) $K^*(892)^{0}$ in $B^\pm \rightarrow K^\pm \pi^+ \pi^-$, (d) $K^*(892)^{0}$ in $B^\pm \rightarrow \pi^\pm K^+ K^-$, and (e) $\phi(1020)$ in $B^\pm \rightarrow K^\pm K^+ K^-$}
    \label{fig:parabolas}
\end{figure}

The efficiency-corrected yields of $B^\pm$ as a
function of $s_\perp$ can be seen in Fig. \ref{fig:parabolas}, with the results of the polynomial function from Eq. \ref{eq:polynomial}. The impact of a \textit{CP} conservation is evident in almost all of the plots, where the vector parabolas of $B^+$ and $B^-$ are very close. The $B^\pm\rightarrow \rho(770)^0 K^\pm$ channel stands out from the others as the only one which shows a significant CP asymmetry, measured to be $A_{CP} = +0.150 \pm 0.019$ with a statistical-only significance of 7.9 standard deviations. For the other channels, the results are: $A_{CP}(\rho(770)^0 \pi^\pm) = -0.004 \pm 0.017$, $A_{CP}(K^{*}(892)^0 \pi^\pm) = -0.015 \pm 0.021$, $A_{CP}(K^{*}(892)^0 K^\pm) = 0.007 \pm 0.054$, $A_{CP}(\phi(1020) K^\pm) = 0.004 \pm 0.014$.

\section{Summary}

Two interesting analyses are presented in this document. The first one showed the inclusive \textit{CP} asymmetries of the
$B^\pm \rightarrow K^\pm \pi^+ \pi^- $, $B^\pm \rightarrow \pi^\pm K^+ K^- $, $B^\pm \rightarrow K^\pm K^+ K^- $, and
$B^\pm \rightarrow \pi^\pm \pi^+ \pi^- $ charmless three-body decays. Significant \textit{CP} asymmetries were found for the latter three \textit{B} decay channels, the last two being observed for the first time. The localized CP asymmetry observed in specific regions of the phase space indicates that \textit{CP} asymmetries are unevenly distributed within it, with both positive and negative $A_{CP}$ values present in the same charged \textit{B} decay channel.

The second analysis employs a novel method to measure \textit{CP} asymmetries in charmless $B\rightarrow PV$ decays, without the
need for amplitude analyses, considering two-body interaction with one spectator meson. It assumes constant magnitudes and phases of amplitudes across the entire phase space, a common assumption in amplitude analyses, supported by the quality of the fits. Five decays are studied and for the $B^\pm\rightarrow \rho(770)^0 K^\pm$ the \textit{CP} asymmetry obtained is $A_{CP} = +0.150 \pm 0.019 \pm 0.011$. For the remaining channels, the observed \textit{CP} asymmetries are compatible with zero.


\bibliographystyle{unsrt} 
\bibliography{proceedings} 



\end{document}